# Identification of Demand through Statistical Distribution Modeling for Improved Demand Forecasting

Murphy Choy
Michelle L.F. Cheong
*School of Information Systems, Singapore Management University,*
*80, Stamford Road, Singapore 178902*
email: murphychoy@smu.edu.sg, michcheong@smu.edu.sg

**Abstract**

Demand functions for goods are generally cyclical in nature with characteristics such as trend or stochasticity. Most existing demand forecasting techniques in literature are designed to manage and forecast this type of demand functions. However, if the demand function is lumpy in nature, then the general demand forecasting techniques may fail given the unusual characteristics of the function. Proper identification of the underlying demand function and using the most appropriate forecasting technique becomes critical. In this paper, we will attempt to explore the key characteristics of the different types of demand function and relate them to known statistical distributions. By fitting statistical distributions to actual past demand data, we are then able to identify the correct demand functions, so that the the most appropriate forecasting technique can be applied to obtain improved forecasting results. We applied the methodology to a real case study to show the reduction in forecasting errors obtained.

## 1.0 Introduction

Demand forecasting is an important aspect of business operation. It is applicable to many different functional areas such as sales, marketing and inventory management. Proper demand forecasting also allows for more efficient and responsive business planning. Because of the benefits it can bring, many industries have paid great attention to demand variability management and forecasting. Tourism and manufacturing are the two major industries who adopt a wide range of demand forecasting and variability management solutions.

There are many factors that affect demand variability. These fluctuations can be attributed to external factors such as changes in trends (rapid change in consumer preference) or events affecting that geographical region (such as major earthquakes or natural disasters, major sports games). Occasionally, fluctuations may also be due to marketing efforts which has successfully piqued the consumer's interest in the products. The supply structure in the economy can also affect the nature of the demand for a good.

There are huge amounts of literature dedicated to demand forecasting as well as demand variability management. Most demand forecasting techniques discussed in the existing literature assumes that the demand function is cyclical in nature with trend. The time varying nature of some demand functions also increased the difficulty in establishing the demand function type and the right model to be used. Lumpy demand function also creates a variety of forecasting problems which are difficult to model using common forecasting techniques.

In this paper, we will describe three types of demand functions and their mathematical representations. We will then simulate demand data using the mathematical representations and model the simulated data to identify the statistical distributions. As such, we would have established the relationship between demand type and statistical distribution of demand data. With actual demand data, we map it to the statistical distribution to identify the demand function. Our proposed methodology is represented in Figure 4 below. This research is motivated by the need to reduce forecasting errors due to the wrongful application of forecasting models without proper identification of the demand function. At the same time, this paper also seeks to demonstrate the reduction in forecasting error for different demand functions using the appropriate forecasting technique.

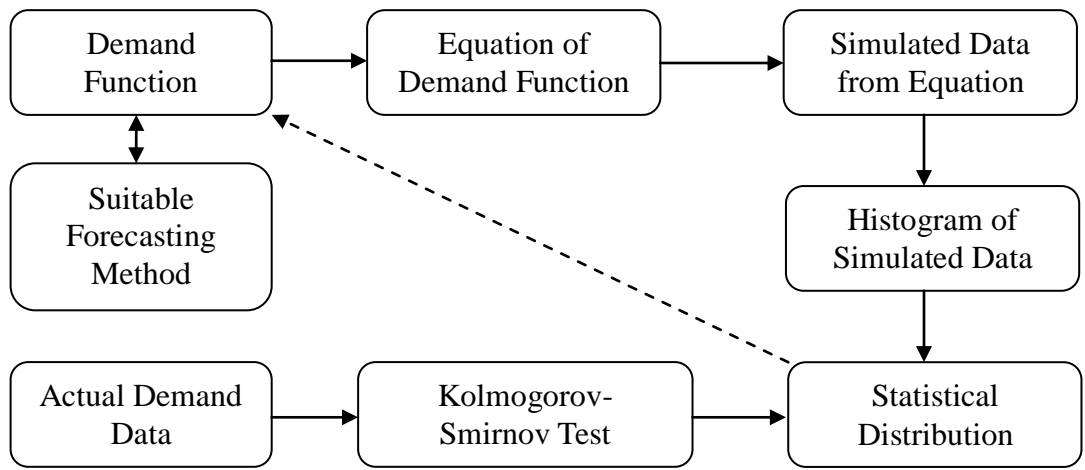

Figure 1: Proposed Methodology

In Section 2.0, we define three different types of demand functions and their respective mathematical representations. In Section 3.0, we review existing literatures on the identifications of the various types of demand functions and the various criteria used to examine them. In Section 4.0, we perform data simulation to relate statistical distribution to demand function. We applied our methodology in Section 5.0 on a real case with demand data which are mapped to different statistical distribution using Kolmogorov-Smirnov Goodness of Fit test and then determine the demand functions. With the identified demand functions, different forecasting methods are applied and their respective forecasting errors are tabulated.

**2.0    Types of Demand Functions**

**TYPE 1** – The first type of demand function is the generic cyclical model with trend. This type of demand function can be generalized into the following form as Equation (1).

Let,

- $Y_i$ = demand of product at time i
- $T_i$ = upward or downward trend component of demand at time i
- $C_{ij}$ = cyclical component of type j at time i, where j = 1 to J

$$Y_i = T_i + C_{i1} + C_{i2} + ... + C_{iJ} + e_i \qquad (1)$$

Cyclical demand functions can be identified easily as their mathematical forms denotes different types of cyclical components such as seasonal trends, periodic trends and other trends which contributes to the overall demand function.

**TYPE 2** – The second type of demand function is commonly known as stochastic demand. Stochastic demand can be considered to be random values where the starting value is derived from previous value or values. It often occurs as time series which is serially correlated. Thus, this type of demand function can be generalized into the following mathematical form as Equation (2).

Let $Y_i$ = demand level at time i

Then, $Y_i = F(Y_{i-1}) + e_i$ \qquad (2)

**TYPE 3** – The last type of demand function is the lumpy demand function. This type of demand resembles stochastic processes but has its own unique characteristics. In the literature on lumpy demand forecasting (Bartezzaghi et al, 1999; Wemmerlöv and Why-bark, 1984), there were several different types of lumpy demand functions discussed. Three main characteristics were summarized from the literature.

- **Variable**: Fluctuations are present and related to some common factors (Wemmerlöv, 1986; Ho, 1995; Syntetos and Boylan, 2005).
- **Sporadic**: Demand can be non-existent for many periods in history (Ward, 1978; Williams, 1982; Fildes and Beard, 1992; Vereecke and Verstraeten, 1994; Syntetos and Boylan, 2005)
- **Nervous**: Each successive observations is different which implies low cross time correlation (Wemmerlöv and Whybark,1984; Ho, 1995; Bartezzaghi and Verganti, 1995).

A lumpy demand distribution is defined as a demand which is extremely irregular with high level of volatility coupled with extensive periods of zero demand (Gutierrez, 2004). While there are several other versions of this definition, all definitions essentially retained the three key characteristics mentioned above. However, there is a fundamental problem with definition of the lumpy distribution.

The sporadic characteristic suggests that demand will be zero for many periods in history. However, one could still have a lumpy demand if there exists a fixed base level of demand greater than zero. Below is a demonstration of the equivalence of the two.

Let,
- $Y_i$ = the level of a normal lumpy demand for time i
- $Z_i$ = the level of a modified lumpy demand for time i
- $F(i)$ = the distribution which created the lumpy demand at time i
- $F'(i)$ = distribution $F(i)$ shifted by a constant value A
- A = fixed base level demand > 0

$$Y_i = F(I) \text{ at } I = i \tag{3}$$

Adding a fixed base level demand A to Yi to get Zi,

$$Z_i = Y_i + A \text{ at } I = i \tag{4}$$

Thus, $Z_i = F'(I)$ at $I = i$ (5)

Given the equivalence in terms of form for (3) and (5), there are no mathematical reasons to exclude any potential lumpy demand function with a fixed base level demand > 0, from the family. So far, we have not found any literature which has a specific mathematical form to explain the phenomena.

Unlike stochastic demand function, lumpy demand function can be identified as a compound distribution between a fixed base level demand and a positive demand function which is usually defined as Geometric distribution or Exponential distribution, represented by F(i) in Equation (3). After mathematically defining the three demand functions that are commonly encountered, let us examine existing literatures on the identifications of the various types of demand functions and the various criteria used to examine them.

**3.0    Literature Review**

Current literatures in this area typically focused on the forecasting solution given a particular type of demand function. There are many papers which talk about various techniques in managing the level of uncertainty (Bartezzaghi et. Al., 1995; Bartezzaghi et. Al., 1999; Syntetos et. Al., 2005). Such papers focused on the development of single algorithm or framework and attempted to measure the performance of such framework against existing ones (Fliedner, 1999).

The second group of papers focused on solving the problems of intermittent demand or lumpy demand with a variety of tools (Ward, 1978; Wemmerlöv et. Al., 1984; Wemmerlöv, 1986) and suggested framework (Vereecke et. Al., 1984, 1994). While some of the papers (Syntetos, 2001) are focused on the problems of the techniques employed, they are still describing the issues given a fixed context. The papers do not generally discuss any of the points about the proper identification of the demand functions while some have mentioned the importance of identification and description of the demand function (Rafael, 2002).

The last group of papers focus on the system that oversees the operation and how improvement to these systems does help (Fildes, 1992). While there are papers describing how the processes are affected by the lumpy demand (Ho, 1995), they again do not attempt to classify the type of demand with respect to the individual characteristics.

In our paper, we will attempt to explore the key characteristics of the different types of demand function and relate them to known statistical distributions. By fitting statistical distributions to actual past demand data, we are then able to identify the correct demand functions, so that the the most appropriate forecasting technique can be applied to obtain improved forecasting results.

## 4.0    Statistical Distribution of Simulated Demand Data

From the mathematical formulations in Section 2.0, we can simulate some data that best represent each group of demand function. It is important to note that the simulations made use of random number generators to demonstrate the distribution of the demand data for a given demand function, so the absolute values are immaterial.

For TYPE 1 – generic cyclical model with trend, we can represent the model in Figure 2.

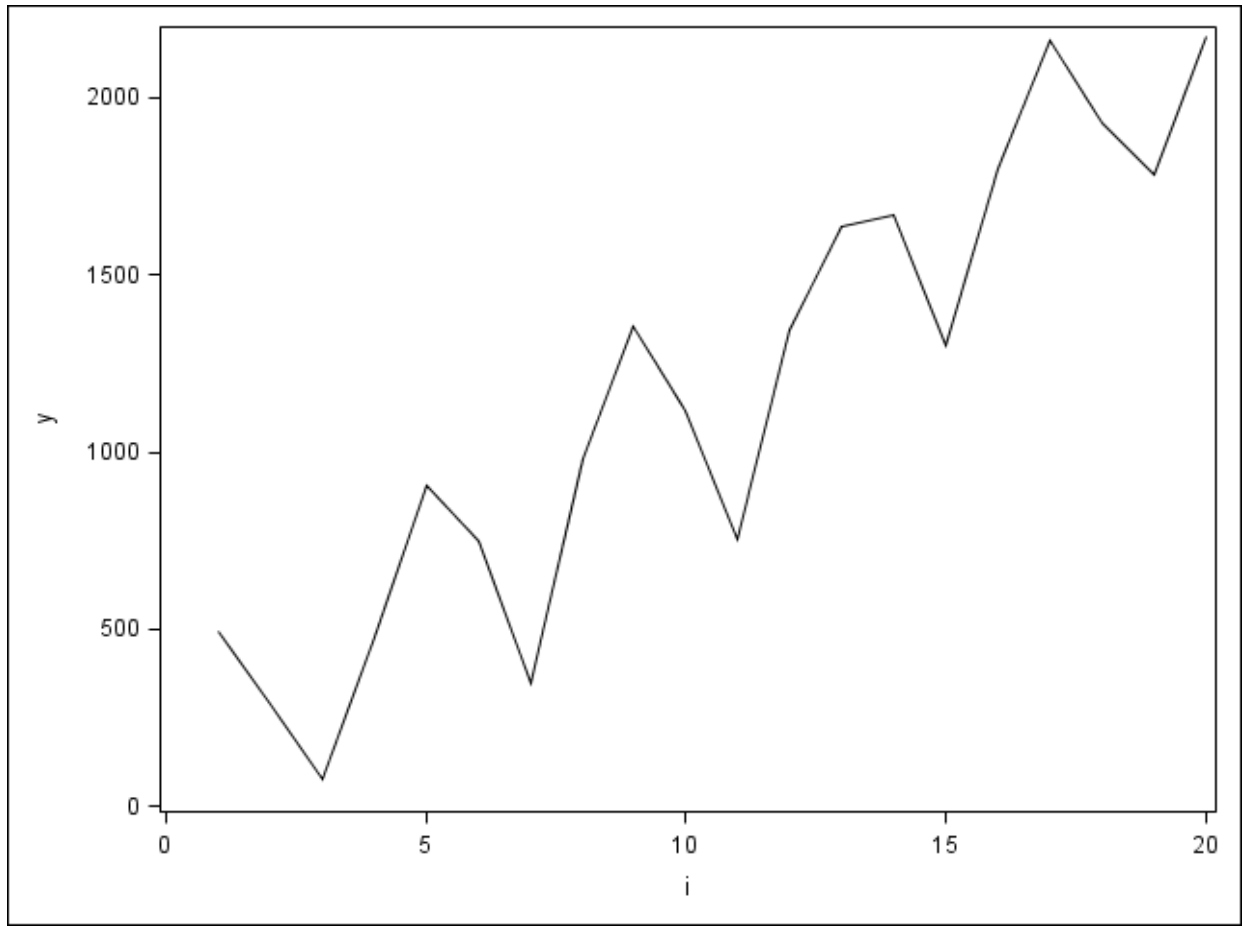

Figure 2: Generic Cyclical Model without Trend

From Equation (1) and Figure 2, we can observe that the cyclical form contains elements from trend and seasonal influence. Given the trends are sinusoidal or regular in nature, which essentially places a constraint on the possible values for the demand data.

Let,
- $C_{ij}$ be the type j cyclical component of the demand at time i
- $A = Min_i \, (\Sigma_j C_{ij})$      for all i

- $B = \text{Max}_i(\Sigma_j C_{ij})$    for all i

Thus,

$A \leq Y_{ij} \leq B$   for j = 1 to J

Since from Equation (1),

$Y_i = T_i + C_{i1} + C_{i2} + \ldots + C_{iJ} + e_i$

Therefore,

$T_i + A \leq Y_i \leq T_i + B$

For a given time i,

$P(Y_i) = 1 / (T_i + B - T_i - A)$

$P(Y_i) = 1 / (B - A)$ which is a uniform distribution

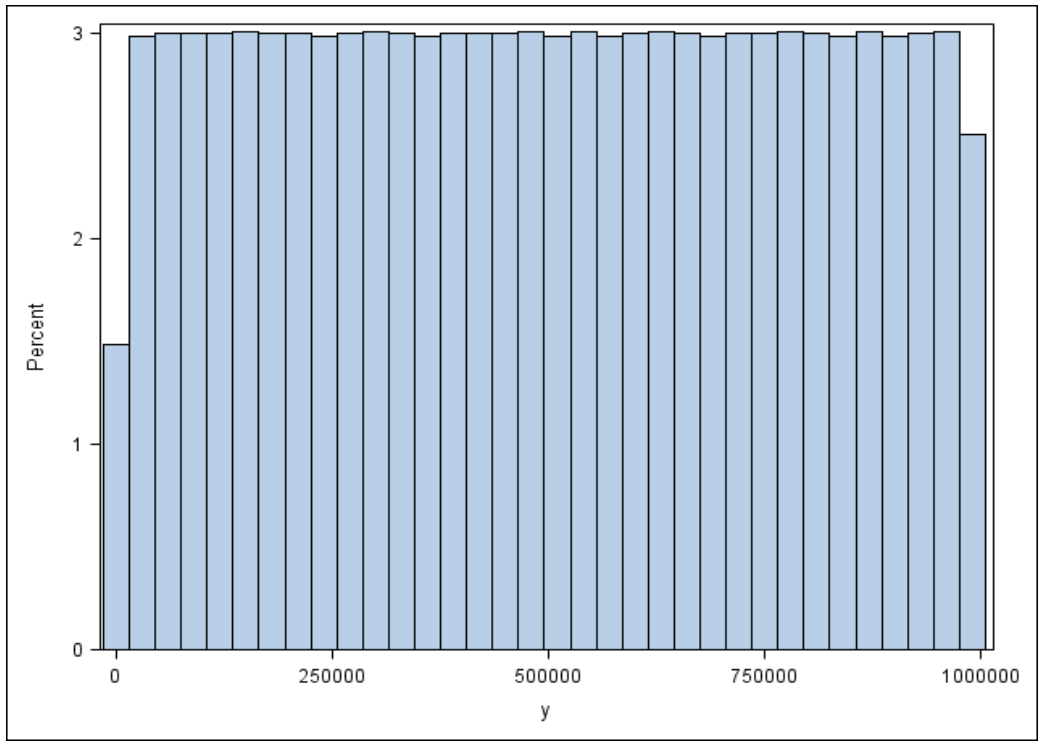

Chart 1: Histogram of Simulated Data from a Cyclic Data Form

From Chart 1, we can see that the histogram from a simulated strong cyclical dataset demonstrated the overall balance of the cycle indicating that demand data seems to be relatively well distributed and appears to be uniform distribution. In essence, we can determine that a particular demand function is cyclical if the histogram of the dataset fits a uniform distribution. The explanation for this is the following. For a cyclical demand function, for each time i, i + S, i + 2S, and so on, where S is the time period for 1 complete cycle, the value of the demand $Y_i$ should be the same value, without considering the trend component. If this characteristic is applied to all $Y_i$ for all i, then each $Y_i$ will occur with equal probability.

For TYPE 2 – stochastic demand, we can represent the model in Figure 3.

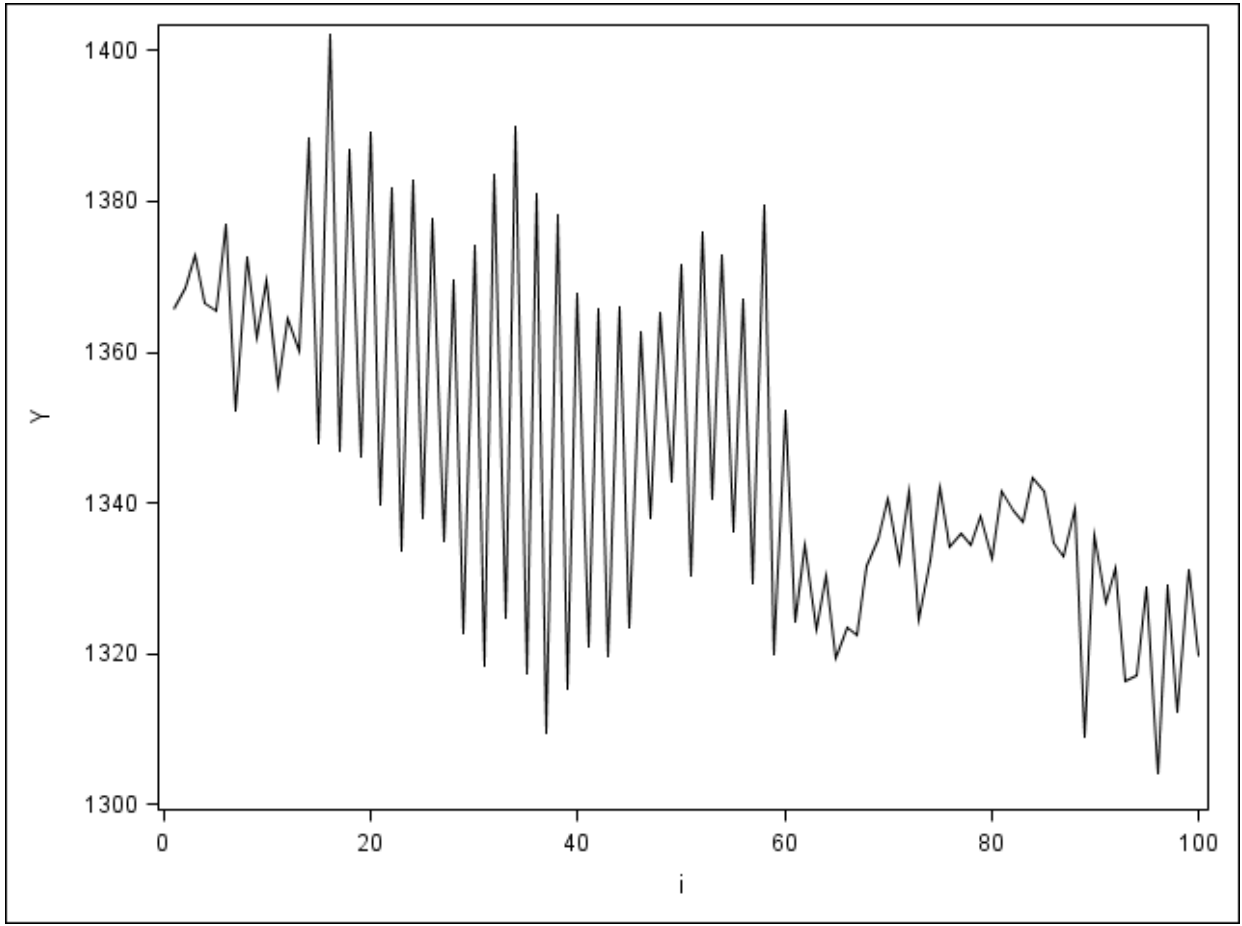

Figure 3: Stochastic Demand Model

From Equation (2) and Figure 3, we can observe that the stochastic demand contains points which have higher frequency around the mean and lower frequency further away from the mean.

To represent a time series, we assign a modifier α in Equation (2) such that,
$$Y_i = F(Y_{i-1}) + e_i = \alpha Y_{i-1} + e_i \qquad (2a)$$

We can expand Equation (2a) as follow,
$$Y_i = \alpha Y_{i-1} + e_i = \alpha (\alpha Y_{i-2}) + e_i = \alpha (\alpha (\alpha Y_{i-3})) + e_i = ... = \alpha^{i-1}(Y_1) + e_i$$
For $0 \leq \alpha \leq 1$, when $i \to \infty$, $\alpha^{i-1} \to 0$, and $Y_i \to e_i$

Thus, the distribution of $Y_i$ effectively follows that of the error term which is essentially a normal distribution.

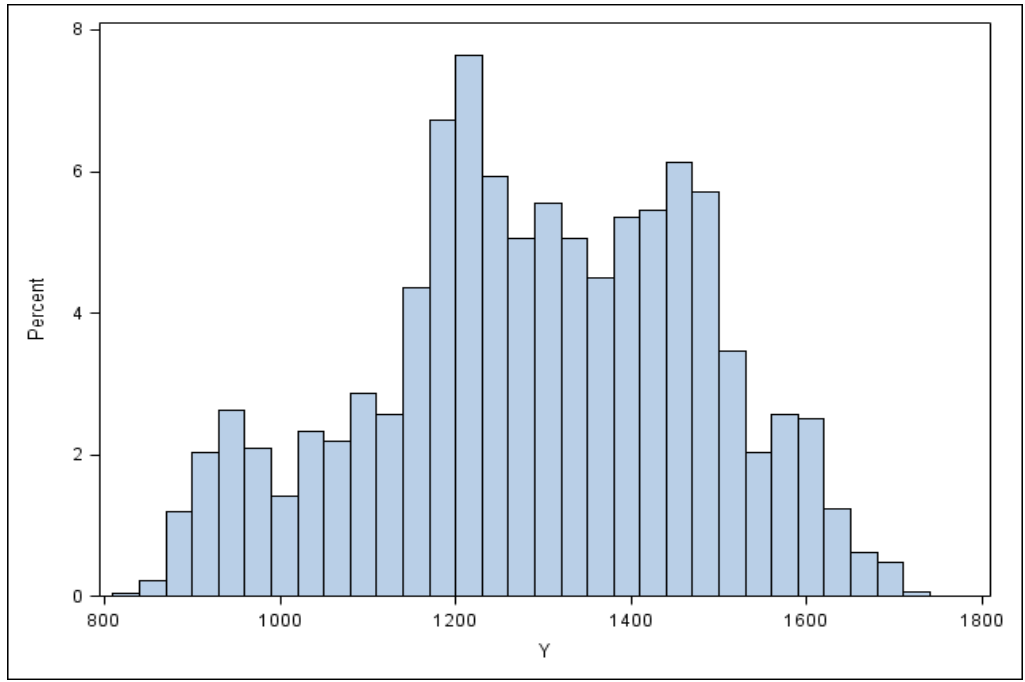

Chart 2: Histogram of Simulated Data from a Stochastic Data Form

From Chart 2, we can see that the histogram of a simulated stochastic dataset is almost normally distributed. This is distinctly different from the histogram of a cyclical demand function which is uniformly distribution. Thus, we can determine that a particular demand function is stochastic if the histogram of the dataset fits a normal distribution. One key reason for this distribution is that the mathematical form of the demand function would dictate that the behavior of the model is more akin to the exponential weighted moving average which fluctuates around a certain average value. At the same time, due to the nature of such process being dependent on the previous observation, the observations can have high serial correlations.

For TYPE 3 – lumpy demand, we can represent the model in Figure 4.

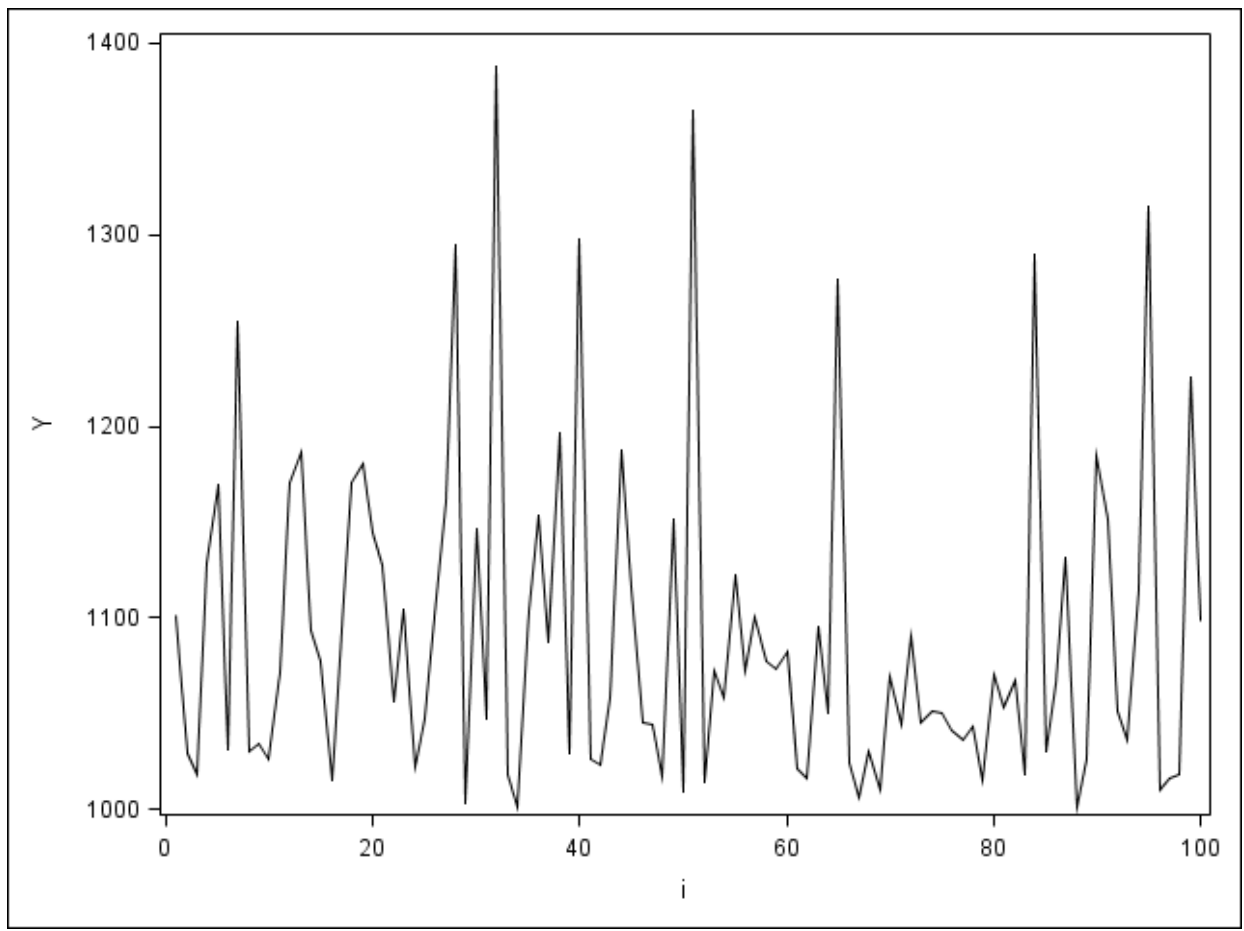

Figure 4: Lumpy Demand Model

As defined in Section 2.0, lumpy demand is a a compounded demand of a statistical distribution and a fixed base level. As the base level is fixed, the distribution of the demand data will follow the distribution that is generating the demand level above the fixed level. In Chart 3 below, we can see the effect of an exponential distribution on the demand data. Thus, we can determine that a particular demand function is lumpy if the histogram of the dataset fits a lumpy stochastic process compounded with a base level demand $\geq 0$.

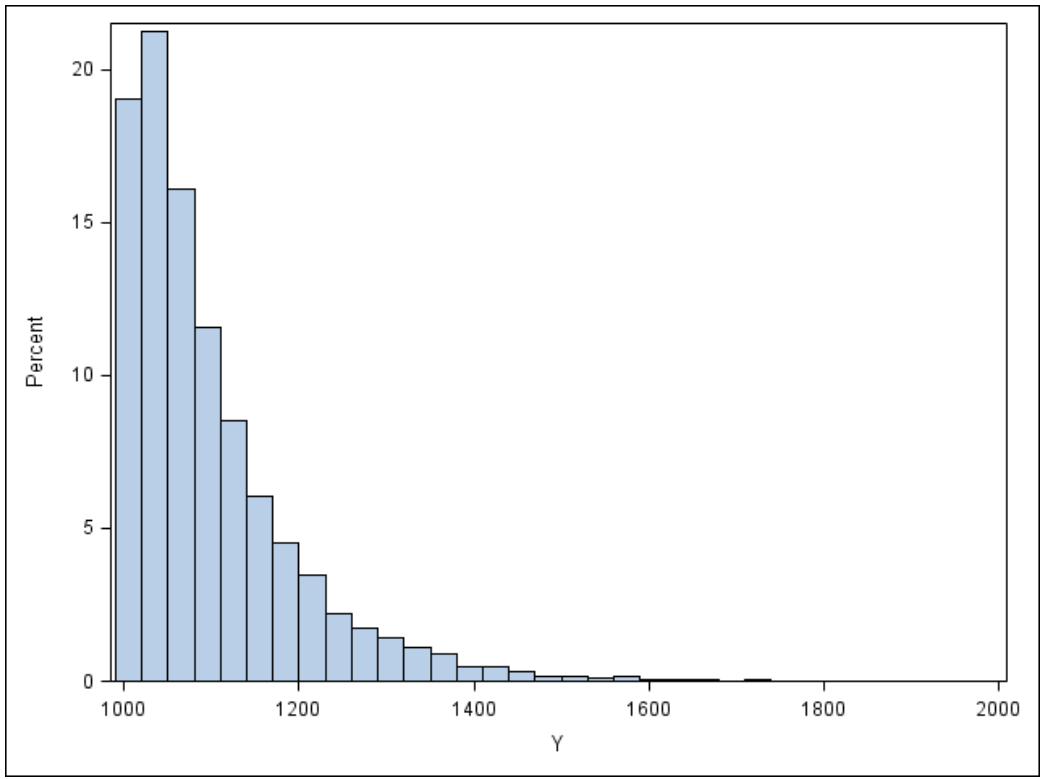

Chart 3: Histogram of Simulated Data from an Exponential Data Form Compounded with a Non-Zero Base Level Demand of 1000

## 5.0     Application of Methodology on Real Case

We apply our methodology on a real case with demand data from a retailer specializing in luxury watches. The retailer is currently facing problems with forecasting the demand for luxury watches in several countries. The high level of volatility in growing economies has thrown most of the forecasts off by a huge error. Currently, the practice is to use a simple moving average in conjunction with manual adjustment. However, this approach only works for demands which are stochastic in nature.

We fitted the past demand data from each country to statistical distributions using the Kolmogorov-Smirnov goodness of fit test. Table 1 below shows the best fitted distributions for each country and their corresponding identified demand type.

| Country | Distribution | Demand Type |
|---------|--------------|-------------|
| A | Exponential | Lumpy |
| B | Normal | Stochastic |
| C | Uniform | Cyclical |
| D | Exponential | Lumpy |
| E | Normal | Stochastic |
| F | Normal | Stochastic |
| G | Normal | Stochastic |
| H | Exponential | Lumpy |
| I | Exponential | Lumpy |

Table 1: Luxury Watches Demand Classification for 9 Countries

From Table 1, we observed that four out of nine demand functions are lumpy demand which supported the reason for large demand forecasting errors due to the use of inappropriate forecasting technique. We applied four different time series forecasting methods to a selected demand of each demand type and calculated the average Mean Squared Error in Table 2.

|  | Demand Type | | |
|---|---|---|---|
| **Forecasting Method** | **Cyclical** | **Stochastic** | **Lumpy** |
| • Exponential Smoothing | 28283664 | 13491035 | 117724405 |
| • Holts Winter (Multiplicative) | 3651334 | 207854 | 59234 |
| • Holts Winter (Additive) | 4336168 | 195455 | **32448** |
| • Stepwise Auto-Regressive | **381348** | **130929** | 56889 |

Table 2: Mean Squared Error of Different Forecasting Methods Applied to Different Demand Types

From Table 2, we can observe that Lumpy demand is best predicted by Holts Winter Additive model as opposed to Stepwise Auto-Regressive model with the lowest average MSE. In fact, the drop in MSE corresponded to approximately 43% improvement in the accuracy of the forecast.

# 6.0 Conclusion

We have attempted to characterize three types of demand functions and using simulated data to establish the relationship between demand function and statistical distributions. We applied the relationship to real demand data so as to correctly identify the demand type and to determine the most appropriate forecasting method with the smallest Mean Squared Error. We have illustrated that a good forecast does not depend solely on the forecasting technique, but also on the correct identification of demand function. At the same time, the paper provided a simple approach to classifying demand functions which does not require complex calculations or evaluation criteria.

uncertainty: A comparative analysis of no-monetary performance variables, International Journal of Production Research, 24 (2), 343-358.

Williams T.M., 1982, Reorder levels for lumpy demand, Journal of the Operational Research Society, 33, 185-189.
14